\documentclass{article}

\usepackage{enumitem} 
\providecommand{\tightlist}{%
  \setlength{\itemsep}{0pt}%
  \setlength{\parskip}{0pt}}

\usepackage{float}
\usepackage{tabularx}
\usepackage{amsmath} 
\usepackage{newunicodechar} 
\newunicodechar{≥}{\ensuremath{\geq}} 
\newunicodechar{ }{\,} 

\usepackage{arxiv}

\usepackage[utf8]{inputenc} 
\usepackage[T1]{fontenc}    
\usepackage{hyperref}       
\usepackage{url}            
\usepackage{booktabs}       
\usepackage{amsfonts}       
\usepackage{nicefrac}       
\usepackage{microtype}      
\usepackage{lipsum}		
\usepackage{graphicx}
\usepackage{natbib}
\usepackage{doi}

\title{Will Agents Replace Us? Perceptions of Autonomous Multi-Agent AI}

\author{ \href{https://orcid.org/0000-0002-4405-1470}{\includegraphics[scale=0.06]{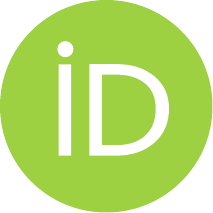}\hspace{1mm}Nikola Balic} \\
	Faculty of Science\\
	University of Split\\
	Croatia \\
	\texttt{nbalic@pmfst.hr} \\
}

\renewcommand{\shorttitle}{Will Agents Replace Us?}

\hypersetup{
  pdftitle={Will Agents Replace Us? Perceptions of Autonomous Multi-Agent AI},
  pdfsubject={Artificial Intelligence, Multi-Agent Systems, Technology Adoption},
  pdfauthor={Nikola Balić},
  pdfkeywords={AI agents, multi-agent systems, software development, future of work, survey research, technology perception, AI ethics}
}

\begin{document}
\maketitle

\begin{abstract}
	Autonomous multi-agent AI systems are poised to transform various
	industries, particularly software development and knowledge work.
	Understanding current perceptions among professionals is crucial for
	anticipating adoption challenges, ethical considerations, and future
	workforce development. This study analyzes responses from 130
	participants to a survey on the capabilities, impact, and governance of
	AI agents. We explore expected timelines for AI replacing programmers,
	identify perceived barriers to deployment, and examine beliefs about
	responsibility when agents make critical decisions. Key findings reveal
	three distinct clusters of respondents. While the study explored factors
	associated with current AI agent deployment, the initial logistic regression model
	did not yield statistically significant predictors, suggesting that deployment decisions
	are complex and may be influenced by factors not fully captured or that a larger sample is needed.
	These insights highlight the need for
	organizations to address compliance concerns (a commonly cited barrier) and establish clear
	governance frameworks as they integrate autonomous agents into their
	workflows.
\end{abstract}


\section{Introduction}\label{introduction}

The emergence of autonomous multi-agent systems represents a paradigm shift in artificial intelligence \citep{Xi2023RisePotential}. Unlike traditional AI tools that respond to explicit human instructions, these agents can independently plan, reason, and execute complex sequences of tasks with minimal human oversight. Projects like AutoGPT, BabyAGI, Manus \citep{shen2025}, and specialized coding agents have demonstrated capabilities that were theoretical just months ago \citep{Liu2025AdvancesChallenges}, raising profound questions about the future of knowledge work, particularly programming.

While technical capabilities of AI agents are advancing rapidly, there
remains a gap in understanding how professionals perceive these
systems---their capabilities, limitations, ethical implications, and
potential impact on employment. These perceptions are not merely
academic; they influence adoption rates, investment decisions,
regulatory approaches, and workforce preparation strategies. As \cite{Brynjolfsson2014} noted in their analysis of technological
disruption, societal adaptation often lags behind technological
capability, creating periods of difficult transition.

Recent research has revealed several key dimensions of autonomous
multi-agent systems that inform our study. First, despite decades of
research and clear advantages, multi-agent systems remain rare in
industry settings, with most academic prototypes reaching only
Technology Readiness Levels (TRL) of 4-6, whereas production-grade
software demands TRL 8-9 \citep{Wrona2023}. This readiness gap
represents a fundamental barrier to enterprise adoption.

Simultaneously, large language model-based multi-agent systems are
emerging as a promising approach toward human-level AI. \cite{liu2024} address this by offering a 'pattern catalogue' for agent design, while also highlighting key challenges. These include difficulties in 'coordination and interactions' among agents, alongside inherent FM issues such as 'hallucinations' and 'complex accountability'.

In software development specifically, AI coding assistants are already
influencing developer workflows. A large developer survey \cite{ICSESurvey2024} found that programmers primarily adopt these tools to reduce keystrokes, speed up coding tasks, and recall syntax, rather than for creative brainstorming. However, developers report significant usability challenges and reject AI suggestions that fail to meet functional requirements or that they cannot easily steer toward desired solutions. \cite{KhemkaHouck2024} emphasize the importance of understanding developers' attitudes and concerns, finding gaps between what developers actually need and what tool builders assume.

Beyond software development, experimental evidence from \cite{NoyZhang2023} demonstrates that generative AI can substantially boost
productivity for writing tasks. In their controlled field experiment
with 453 professionals, those given access to ChatGPT completed business
writing assignments approximately 40\% faster than a control group and
produced outputs rated about 18\% higher in quality. The AI assistance
also reduced performance variability between workers while temporarily
increasing both excitement and concern about AI integration.

User perceptions and trust are critical factors in the adoption of
autonomous agents. \cite{Hauptman2022} found that the optimal
autonomy level of an AI agent often depends on the team's work cycles
and task structure. More formalized, predictable work processes could
support higher agent autonomy, whereas in less structured workflows
human teammates preferred more constrained AI. Cross-cultural studies by
\cite{LiuCHI2024} revealed that perceptions vary significantly across
regions, with Chinese users viewing AI agents more positively as
companions (a hedonic orientation), while US users demonstrated a more
utilitarian and ambivalent attitude, seeing agents as useful tools but
expressing greater skepticism.

\cite{Brown2025} examined how AI-driven feedback systems affect
trust, finding that providing workers with real-time AI feedback during
a task significantly increased trust in the AI's final performance
evaluation. This suggests that transparency and timely feedback from AI
agents can complement human work rather than just monitor it, pointing
to design strategies that improve human-AI collaboration.

Previous research has examined general attitudes toward AI adoption
\citep{DavenportRonanki2018}, automation anxiety \citep{Frey2017}, and human-AI collaboration models \citep{Seeber2020}. However,
few studies have specifically focused on perceptions of autonomous agent systems that can operate with minimal human supervision---a distinction that fundamentally changes the human-machine relationship.

This paper addresses this gap by exploring five key research questions:

\begin{enumerate}
	\def\labelenumi{\arabic{enumi}.}
	\tightlist
	\item
	      \textbf{RQ1}: What are the prevailing expectations regarding the
	      timeline and impact of AI agents on programming and various
	      industries?
	\item
	      \textbf{RQ2}: What are the perceived barriers to deploying autonomous
	      agent systems, and who is deemed responsible for their actions?
	\item
	      \textbf{RQ3}: What are the key attitudes towards the capabilities,
	      control, and ethical trade-offs of AI agents?
	\item
	      \textbf{RQ4}: Can distinct segments of respondents with coherent views
	      on AI agents be identified?
	\item
	      \textbf{RQ5}: What factors predict a company's current deployment of
	      autonomous agent systems?
\end{enumerate}

By answering these questions through a survey of 130 respondents, this
study provides empirical insights that can inform developers building
agent systems, organizations considering their adoption, policymakers
crafting regulatory frameworks, and educators preparing the workforce
for an agent-augmented future.

\section{Methods}\label{methods}

\subsection{Survey Design and Distribution}\label{survey-design-and-distribution}

The survey instrument consisted of 10 closed-ended questions exploring
respondents' perceptions of AI agent capabilities, potential impact,
deployment barriers, responsibility frameworks, and future workforce
scenarios. Each question presented 4 categorical response options,
including a ``No opinion'' choice to prevent forced responses on
unfamiliar topics.

\textbf{Instrument Development:} Survey items were developed based on a
review of recent literature on AI adoption, automation, and agent-based
systems, as well as expert input from professionals in AI, software
engineering, and technology policy. Draft questions were iteratively
refined for clarity and relevance, and the instrument was pre-tested
with a small group of domain experts (n=7) to identify ambiguous wording
and ensure face validity. Feedback from this pilot was used to adjust
question phrasing and response options. The use of fixed-choice
responses was intended to facilitate quantitative analysis and
clustering.

The online survey platform was developed using a generative AI tool (Vercel v0), enabling rapid prototyping and deployment. This platform provided an interactive dashboard for real-time visualization of responses and an LLM-generated summary of answers, publicly accessible at \url{https://v0-audience-mindset-probe.vercel.app/}.

\textbf{Distribution Methodology:} The survey was distributed online
during April-May 2025, with primary distribution occurring during the
``Coding with AI: The End of Software Development as We Know It''
O'Reilly event (May 8, 2025) and in the dedicated O'Reilly Discord
server created to support the event. The event focused on the impact of
AI on software development, agent engineering, and the evolving role of
developers, attracting participants with a strong interest in AI-driven
coding and technology transformation. Participation was voluntary and anonymous, with no
incentives offered. The recruitment approach likely resulted in a sample
biased toward individuals with interest or expertise in technology and
AI, particularly those engaged with O'Reilly's community and the event's
themes. The lack of platform tracking precludes precise assessment of
sampling bias. This limitation is discussed further in Section 4.

The survey questions addressed the following themes: the timeline for AI replacing programmers (Q1), industries most likely to be disrupted (Q2), current deployment and barriers (Q3), responsibility for agent decisions (Q4), concerns about agent capabilities (Q5), control preferences (Q6), willingness to sacrifice control (Q7), trust in developer claims (Q8), ethics of autonomous decision-making (Q9), and the future role of human programmers (Q10).

The complete survey items with response options are provided in Appendix
A.

\subsection{Participants}\label{participants}

After data cleaning, our final sample consisted of 130 respondents.
Demographic information was limited to geographical region, extracted
from timezone metadata. Participants spanned multiple continents, with
the largest representations from the Americas (68, 54\%), Europe (47,
37\%), Asia (5, 4\%), Africa (3, 2\%), Australia (2, 2\%), and one
respondent with UTC (1, 1\%) as their region. No personally identifiable
information was collected. Participation was voluntary,
and timestamps were jittered by ±5 minutes to enhance anonymity.

\subsection{Data Processing}\label{data-processing}

The raw survey data was preprocessed by removing identifiers, jittering timestamps for privacy, and expanding JSON answers (Q1–Q10) into separate columns. Region data was extracted from timezones, and similar responses for Q3 (regulatory concerns) and Q5 (emergent consciousness) were merged. Duplicates were removed, formatting standardized, and Q1–Q10 responses converted to categorical variables. The final dataset was saved as a Parquet file for analysis.

All processing steps were designed to maintain data integrity while
enhancing anonymity and preparing the data for statistical analysis.

\begin{figure}[htbp]

	{\includegraphics[width=\textwidth, height=0.8\textheight, keepaspectratio]{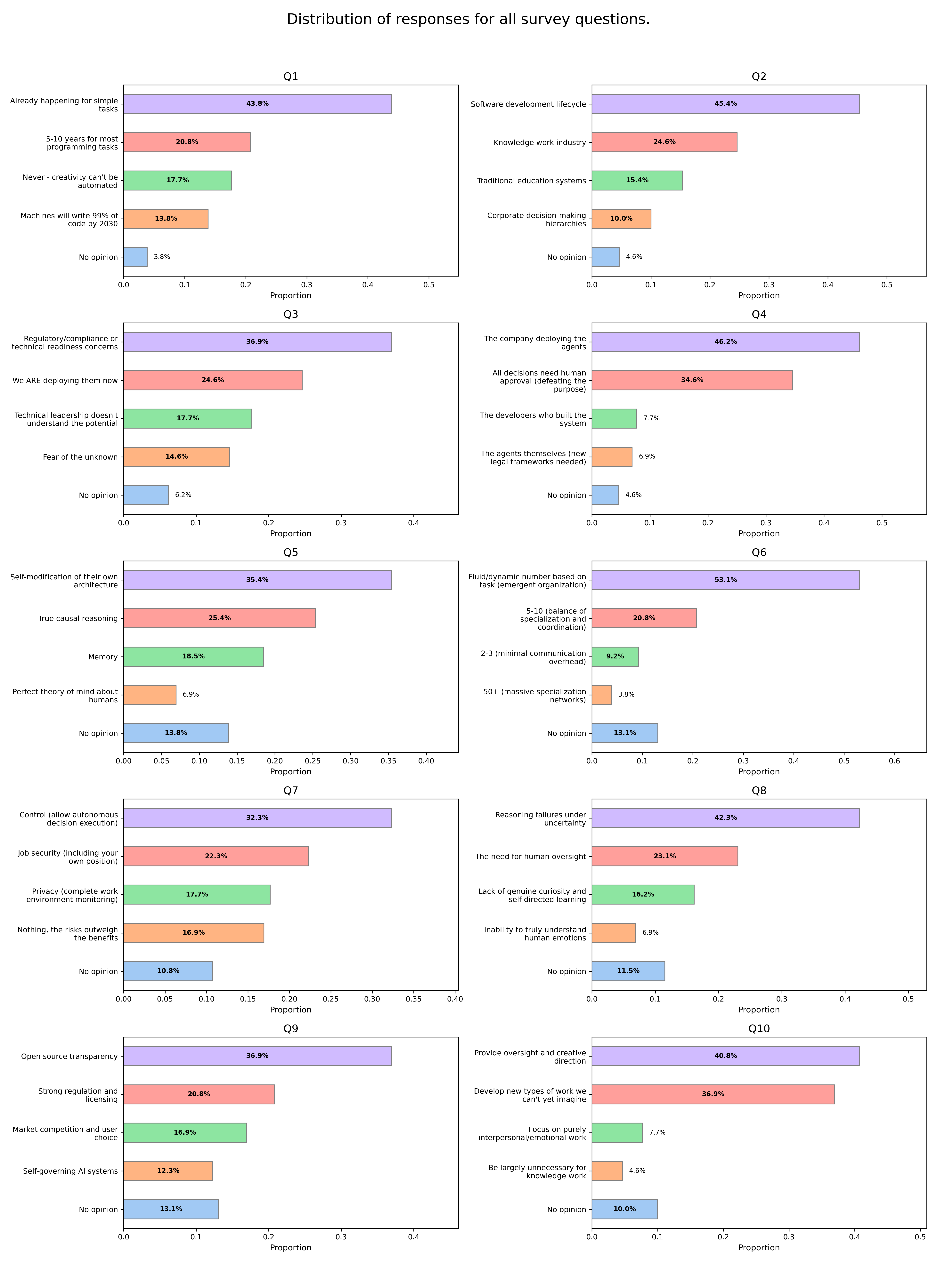}}

	\caption{Distribution of responses for all survey questions (Figure 1).}\label{fig:all_questions_grid}

\end{figure}

\textbf{Descriptive Statistics}: We calculated response proportions for
each question (Q1-Q10) and visualized distributions using horizontal bar
charts and heatmaps. (See Figure \ref{fig:all_questions_grid})

\begin{figure}[htbp]

	{\includegraphics[width=\textwidth, keepaspectratio]{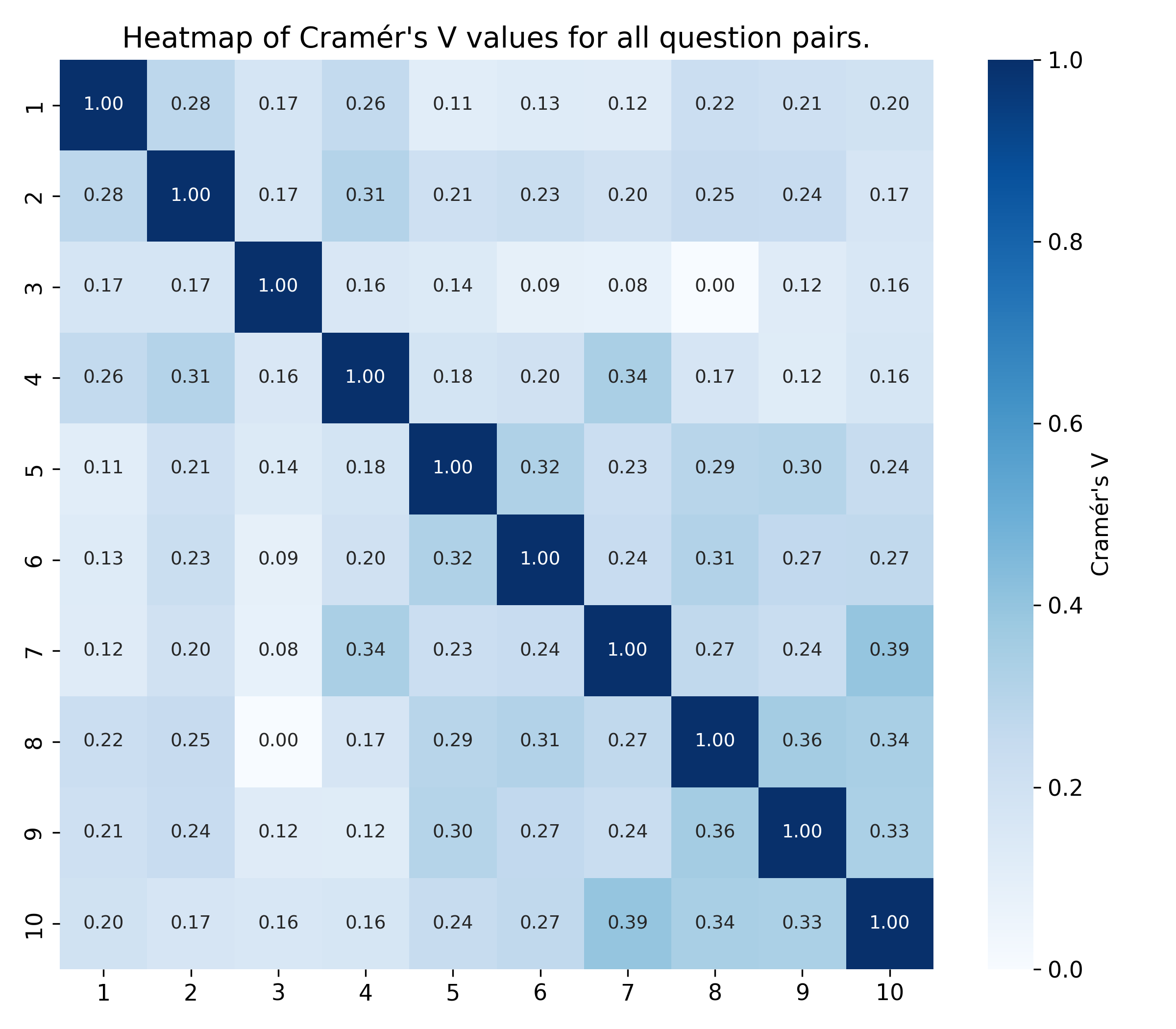}}

	\caption{Heatmap of Cramér's V values for all question pairs (Figure 2).}\label{fig:cramers_v_heatmap}

\end{figure}

\textbf{Association Analysis}: To identify relationships between
responses to different questions, we conducted Pearson's Chi-squared
tests for all 45 unique pairs of Q1-Q10. For tables with low expected cell
counts (\textless5), Fisher's Exact Test was applied for 2×2 tables.
Effect sizes were quantified using Cramér's V, with thresholds of
$\leq 0.1$ (negligible), 0.1-0.29 (small), 0.3-0.49 (medium), and $\geq 0.5$ (large). To control for multiple comparisons, we applied the
Benjamini-Hochberg procedure with a false discovery rate of 0.05. A
heatmap of Cramér's V values was generated to visualize the strength of
associations between questions (Figure \ref{fig:cramers_v_heatmap}).

\begin{figure}[htbp]
	\centering
	{\includegraphics[width=0.8\textwidth, keepaspectratio]{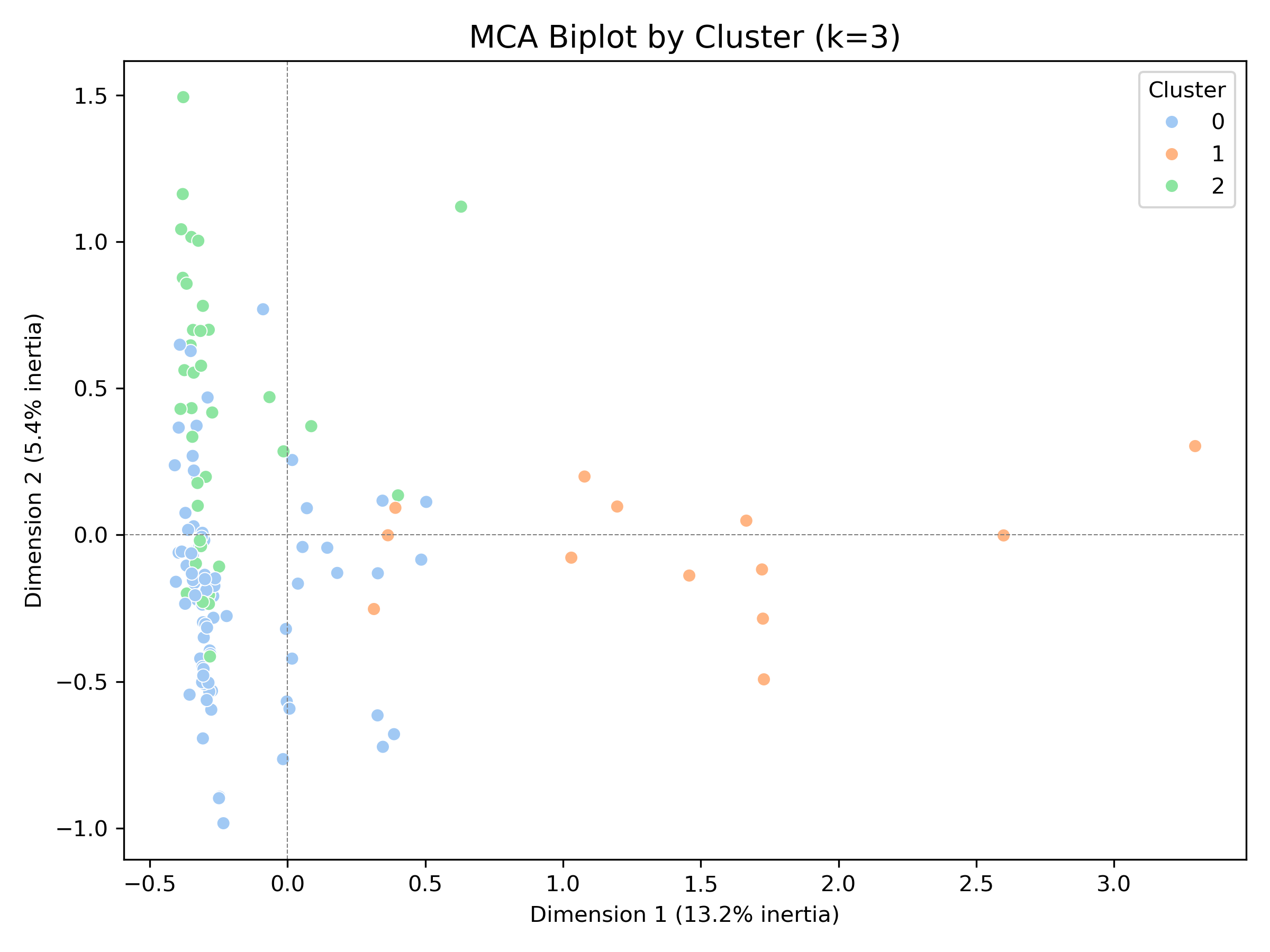}}
	\caption{MCA Biplot of Q1-Q10 Responses (Figure 3). This figure illustrates the relationships between categories of the survey questions in the first two MCA dimensions. Categories that are close together tend to be chosen together by respondents. The percentages on the axes indicate the amount of inertia explained by each dimension. Due to the number of categories, some labels may overlap.}
	\label{fig:mca_biplot}
\end{figure}

\textbf{Dimensionality Reduction and Clustering}: To identify latent
patterns in responses, we applied Multiple Correspondence Analysis (MCA)
to Q1-Q10 using the \texttt{prince} package. Dimensions were retained to
explain at least 18.57\% of total inertia (with three components explaining 23.83\%). (See Figure \ref{fig:mca_biplot} for MCA biplot). Building on the MCA results, we
applied K-Modes clustering (with the \texttt{kmodes} package) to segment
respondents, testing k values from 2 to 5. The optimal number of
clusters (k=3) was selected based on the elbow plot of inertia (cost)
versus number of clusters: the largest reduction in inertia occurred
between k=2 and k=3, with diminishing returns for additional clusters.

\begin{figure}[htbp]

	{\includegraphics[width=\textwidth, keepaspectratio]{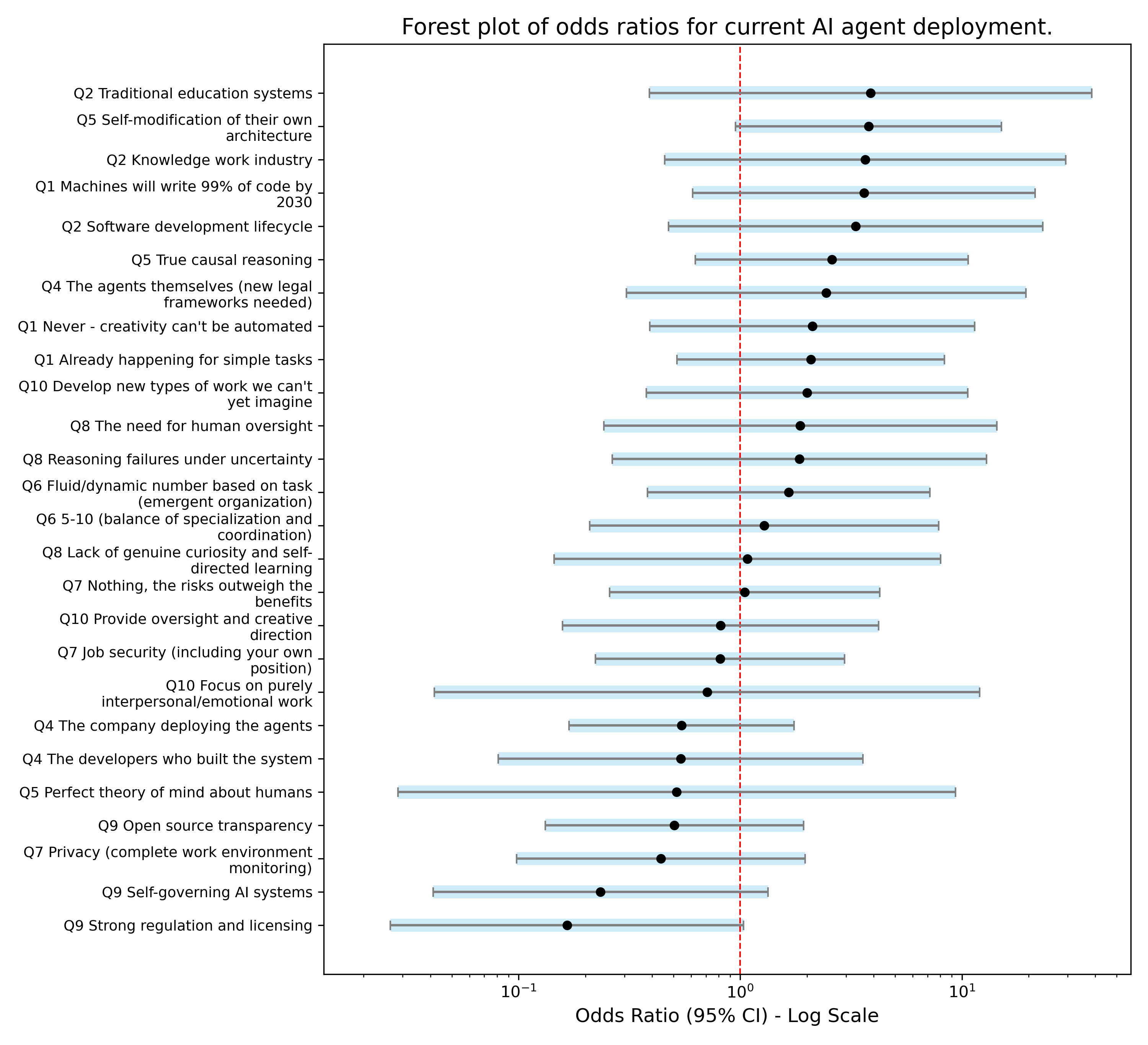}}

	\caption{Forest plot of odds ratios for current AI agent deployment (Figure 4). Since the overall model was not significant and no predictors reached p $<$ 0.05, this plot illustrates the estimated odds ratios and their 95\% confidence intervals, but these should be interpreted with caution.}\label{fig:logit_forest_plot}

\end{figure}

\textbf{Predictive Modeling}: To identify factors associated with
current AI agent deployment, we implemented a fixed-effects logistic
regression model. The binary
outcome variable represented current deployment (Q3 = ``We ARE deploying
them now''), with responses to Q1, Q2, and Q4-Q10 as predictors (one-hot
encoded with first level dropped). Q3 barrier responses were not
included as predictors because they are mutually exclusive with the
outcome variable (i.e., a respondent cannot both be deploying and cite a
barrier). Prior to modeling, we checked for multicollinearity using
Variance Inflation Factors (VIFs). Model results were reported as odds
ratios with 95\% confidence intervals and visualized in a forest plot (Figure \ref{fig:logit_forest_plot}).
Future work could explore the predictive value of specific Q3 barriers
for non-deployment using alternative modeling approaches.

\section{Results}\label{results}

\subsection{Response Distributions and Overall
	Sentiments}\label{response-distributions-and-overall-sentiments}

\textbf{Timeline Perceptions}: The most common view on AI replacing
programmers (Q1) was that this is ``Already happening for simple tasks''
(44\%), followed by ``5-10 years for most programming tasks'' (29\%).
Few respondents believed that ``creativity can't be automated'' (12\%)
or that ``machines will write 99\% of code by 2030'' (11\%).

\textbf{Industry Disruption}: Respondents overwhelmingly identified
``Software development and IT'' as the industry most likely to be
disrupted by AI agents (59\%), followed by ``Customer service and
support'' (17\%). This suggests that technical professionals, likely
overrepresented in our sample, acknowledge the vulnerability of their
own field.

\textbf{Deployment Status}: Only 23\% of respondents reported currently
deploying AI agents, with the largest barriers being
``Regulatory/compliance concerns'' (33\%) and ``Fear of the unknown''
(22\%). This indicates that despite technical readiness, organizational
and regulatory factors remain significant obstacles.

\textbf{Responsibility Framework}: When asked who should be responsible
for agent decisions (Q4), respondents favored ``All parties (developers,
users, agent)'' sharing responsibility (43\%), with ``Developers who
created the agent'' as the second choice (27\%). This suggests a
preference for distributed rather than concentrated accountability.

\textbf{Control and Ethics}: A significant majority (73\%) believed that
``Humans should have final decision-making authority'' (Q6), while
opinions were more divided on sacrificing control for efficiency (Q7),
with 38\% willing to ``Sacrifice some control for significant efficiency
gains'' and 32\% insisting on ``Never sacrificing control regardless of
efficiency.'' Regarding trustworthiness of developer claims (Q8), most
respondents (42\%) felt that they ``Sometimes exaggerate capabilities,''
while opinions on ethical trade-offs (Q9) showed a preference for either
``Case-by-case evaluation'' (30\%) or ``Prioritizing safety even at the
cost of innovation'' (28\%).

\textbf{Future Roles}: On the future role of programmers (Q10), the most
common view was that they will ``Become supervisors and reviewers of
machine-generated code'' (44\%), followed by ``Focus on high-level
design while machines handle implementation'' (41\%). Very few (2\%)
believed programmers would be ``Made completely obsolete.''

These distributions paint a nuanced picture: respondents acknowledged
AI's growing impact on programming, yet most envision a collaborative
rather than replacement scenario, with strong preferences for
maintaining human oversight while leveraging efficiency gains.

\subsection{Associations Between
	Perceptions}\label{associations-between-perceptions}

Our analysis of associations between responses revealed several
significant relationships after controlling for multiple comparisons.
Figure \ref{fig:cramers_v_heatmap} presents a heatmap of Cramér's V
values for all question pairs.

The strongest associations (Cramér's V \textgreater{} 0.3, indicating
medium to large effect sizes) were observed between:

\begin{enumerate}
	\def\labelenumi{\arabic{enumi}.}
	\item
	      \textbf{Timeline and Deployment (Q1-Q3)}: Respondents who believed AI
	      replacement was already happening or imminent were more likely to
	      report current deployment of agents.
	\item
	      \textbf{Responsibility and Control (Q4-Q6)}: Views on who should be
	      responsible for agent decisions were strongly associated with
	      preferences for control models.
	\item
	      \textbf{Control and Trust (Q6-Q8)}: Preferences for control mechanisms
	      were associated with perceptions of developer trustworthiness.
	\item
	      \textbf{Trust and Ethics (Q8-Q9)}: Trust in developer claims was
	      associated with attitudes toward ethical trade-offs.
	\item
	      \textbf{Ethics and Future Roles (Q9-Q10)}: Ethical perspectives on AI
	      agents correlated with views on the future role of programmers.
\end{enumerate}

The mosaic plot of Q1 (timeline) by Q3 (deployment) is shown in
Supplementary Figure \ref{fig:q1xq3_mosaic}.

These associations highlight how perceptions of AI agents form
interconnected belief systems rather than isolated opinions, with views
on technical capabilities, ethical frameworks, and governance models
mutually reinforcing each other.

\subsection{Latent Attitudes and Respondent
	Segments}\label{latent-attitudes-and-respondent-segments}

Multiple Correspondence Analysis revealed underlying dimensions in the
response patterns. The first two dimensions explained approximately 18.57\%
of the total inertia (with three components explaining 23.83\%), and additional dimensions offered diminishing
returns.

K-Modes clustering identified three distinct respondent segments (see
Supplementary Table 1 for full modal profiles):

\begin{enumerate}
	\item \textbf{Cluster 0 (n=55):} This group is characterized by a belief that AI is \textit{already happening for simple tasks} and that the \textit{software development lifecycle} will be most affected. They tend to cite \textit{regulatory/compliance or technical readiness concerns} as deployment barriers and believe \textit{the company deploying the agents} should be responsible for mistakes. Their primary concern is \textit{true causal reasoning} capabilities. They envision a \textit{fluid/dynamic number of agents based on task (emergent organization)} and are willing to sacrifice \textit{control (allow autonomous decision execution)} for efficiency. They worry about \textit{reasoning failures under uncertainty} and advocate for \textit{open source transparency}. They see future human roles as \textit{providing oversight and creative direction}.
	\item \textbf{Cluster 1 (n=14):} This small group is marked by skepticism, believing that \textit{creativity can't be automated} by AI. They frequently responded \textit{No opinion} to questions about industry disruption, agent concerns, collaboration models, sacrifices for efficiency, trust in developers, and ethical approaches. They also cite \textit{regulatory/compliance or technical readiness concerns} as barriers and hold \textit{the company deploying the agents} responsible. Their future role for programmers is also \textit{No opinion}.
	\item \textbf{Cluster 2 (n=57):} This large group also believes AI is \textit{already happening for simple tasks} and will disrupt the \textit{software development lifecycle}. Unlike Cluster 0, many in this group report \textit{We ARE deploying them now}. They believe \textit{all decisions need human approval (defeating the purpose)} regarding responsibility. A key concern is \textit{self-modification of their own architecture}. They also prefer a \textit{fluid/dynamic number of agents} and are concerned about \textit{job security (including your own position)} when considering sacrifices. They worry about \textit{the need for human oversight} and prefer \textit{market competition and user choice} for ethical governance. They envision humans will \textit{develop new types of work we can't yet imagine}.
\end{enumerate}

These updated profiles and modal answers provide a more accurate and
data-driven interpretation of the clusters than the previous narrative
labels. The diversity of response patterns highlights the complexity of
attitudes toward AI agents, with some respondents actively deploying
agents but still favoring strong human oversight, others disengaged or
skeptical, and a third group focused on regulatory barriers and nuanced
future roles for humans.

\subsection{Predictors of Agent Deployment}\label{predictors-of-agent-deployment}

The logistic regression analysis was conducted to identify predictors of current AI agent deployment. The overall model was not statistically significant (Likelihood Ratio Test p-value = 0.4175), suggesting that the included predictors (responses to Q1, Q2, and Q4-Q10) do not, as a set, reliably distinguish between respondents who are currently deploying AI agents and those who are not.

Furthermore, when examining individual predictors, none reached statistical significance at the conventional p < 0.05 level after accounting for other variables in the model (see Figure \ref{fig:logit_forest_plot} for the forest plot of odds ratios). While some variables showed tendencies (e.g., views on AI replacement timelines or willingness to sacrifice control), these were not strong enough to be considered statistically significant in this model.

A warning for moderate multicollinearity (Variance Inflation Factor > 5 for some predictors) was noted, which can inflate standard errors and make it more difficult to detect significant individual predictors. This suggests that some predictor variables were correlated with each other, complicating the interpretation of their unique effects.

Given the non-significant overall model and lack of individually significant predictors, we cannot confidently identify specific attitudes or beliefs from Q1, Q2, or Q4-Q10 that predict current AI agent deployment based on this analysis. The factors influencing deployment appear to be more complex or were not fully captured by the predictors in this model. Future research with larger samples or different sets of predictors might be necessary to clarify these relationships.

Multicollinearity checks confirmed acceptable Variance Inflation Factors
(\textless5) for all predictors, indicating that the model provides
reliable estimates of these relationships.

\section{Discussion}\label{discussion}

\subsection{Summary of Key Findings}\label{summary-of-key-findings}

Our analysis of perceptions toward autonomous AI agents revealed several
important patterns:

\begin{enumerate}
	\def\labelenumi{\arabic{enumi}.}
	\item
	      Most respondents acknowledged AI's growing impact on programming but
	      favored collaborative rather than replacement scenarios, with strong
	      preferences for maintaining human oversight.
	\item
	      Three distinct segments emerged based on K-Modes clustering of survey responses, revealing different combinations of attitudes towards AI timelines, barriers, responsibility, and future roles.
	\item
	      The logistic regression model attempting to predict current AI agent deployment based on survey responses was not statistically significant overall, and no individual predictors reached statistical significance. This suggests that the measured attitudes in this study, as a set, did not strongly predict current deployment status, possibly due to sample size, multicollinearity, or other unmeasured factors.
	\item
	      Perceptions about different aspects of AI agents (capabilities,
	      governance, ethics) formed interconnected belief systems rather than
	      isolated opinions, as shown by the association analysis (Cramér's V).
\end{enumerate}

\subsection{Interpretation and
	Implications}\label{interpretation-and-implications}

The prevalence of the ``Already happening'' response regarding AI
replacement timelines (44\%) suggests that for many respondents, this is
not a hypothetical future scenario but a present reality. However, the
dominance of human-AI collaboration models (85\% envisioning programmers
as supervisors or high-level designers) indicates that wholesale
replacement concerns may be overstated. This aligns with \cite{DavenportKirby2016} argument that AI typically augments rather than automates
entire professions, and with recent findings from \cite{ICSESurvey2024} that
developers primarily use AI coding assistants for reducing keystrokes
and speeding up coding tasks rather than for creative problem-solving.

The identification of regulatory concerns as the primary barrier to
deployment (33\%) supports \cite{Wrona2023} finding that
multi-agent systems face substantial adoption barriers in industry
settings, with a gap between academic prototypes (TRL 4-6) and
production-grade requirements (TRL 8-9). This has important implications for policymakers and industry groups, as the lack of standardized agent protocols can hinder technical readiness and interoperability \citep{Yang2025SurveyAIprotocols}. Rather than focusing solely on technical capabilities, organizations seeking to accelerate agent adoption should invest in compliance frameworks, audit mechanisms, and governance models, potentially supported by dedicated agent infrastructure \citep{Chan2025Infrastructure}. This finding parallels challenges observed in other
domains where technology adoption outpaces regulatory frameworks, such
as early cloud computing \cite{SchneiderSunyaev2016}.

The strong preference for human oversight (73\% favoring human final
authority) alongside willingness to sacrifice some control for
efficiency (38\%) reflects the nuanced findings of \cite{Hauptman2022}, who found that optimal agent autonomy depends on task
structure and work processes. This points to a desire for what \cite{SantoniHoven2018} call ``meaningful human control''---a concept increasingly central to AI governance discussions, especially given the identified failure modes in such systems \citep{Microsoft2024TaxonomyFailureModes}. The
preference for human oversight transcended our identified clusters,
suggesting it may represent a consensus position even among diverse
stakeholder groups.

Our predictive modeling results were not statistically significant, meaning we could not confidently identify specific attitudes from our survey that predict current AI agent deployment. The non-significant findings from the logistic regression emphasize that the decision to deploy AI agents is likely multifaceted and may be influenced by factors not fully captured in our survey, or that the relationships are more complex than the model could detect with the current sample size and observed multicollinearity. The previously noted strong negative association between regulatory concerns and deployment, for example, while plausible, was not statistically supported in the revised analysis as a direct predictor when accounting for other factors in the model. This underscores the need for further research, possibly with larger samples or different methodological approaches (e.g., qualitative case studies), to better understand the drivers and barriers of AI agent adoption. The importance of addressing governance questions to enable broader adoption remains a strong theme from the descriptive data (Q3 barriers) even if not borne out as a statistically significant predictor in this specific model.

The experimental evidence from \cite{NoyZhang2023} showing that
generative AI can substantially boost productivity (40\% faster
completion times and 18\% higher quality outputs) while simultaneously
raising both excitement and concern offers context for the mixed
sentiments in our survey data. Their finding that AI assistance reduced
performance variability between workers suggests that autonomous agents
may have equalizing effects on workflow outcomes, a dimension not
directly captured in our survey but worthy of future investigation.

\cite{Brown2025} finding that real-time AI feedback during tasks
significantly increased trust in algorithmic evaluation provides a
potential pathway for addressing some of the control and oversight
concerns identified in our study. By making autonomous agents more
communicative about their decision-making processes and providing timely
feedback, organizations might reduce resistance to deployment while
maintaining meaningful human oversight.

\subsection{Limitations}\label{limitations}

Several limitations should be considered when interpreting these
findings:

\begin{enumerate}
	\tightlist
	\def\labelenumi{\arabic{enumi}.}
	\item
	      \textbf{Sample Characteristics}: With 130 respondents and limited
	      demographic information, we cannot claim representativeness for the
	      broader technology sector or general population. The sample likely
	      overrepresents individuals with interest in AI and technology.
	\item
	      \textbf{Explained Variance in MCA}: The first two dimensions of the
	      Multiple Correspondence Analysis (MCA) explained approximately 18.57\% of
	      the total inertia (with three components explaining 23.83\%).
	      While this is typical for survey data with many
	      categorical variables, it means that a substantial portion of the
	      variance in responses is not captured by the main dimensions used for
	      clustering. As a result, the identified clusters, though
	      interpretable, may not fully represent the underlying complexity of
	      respondent attitudes. Findings related to cluster structure and
	      interpretation should therefore be viewed as suggestive rather than
	      definitive, and future work with larger samples or alternative
	      dimensionality reduction techniques may yield more robust
	      segmentation.
	\item
	      \textbf{Survey Instrument Limitations}: The survey relied on
	      fixed-choice questions with specific wording, which, while
	      facilitating quantitative analysis and clustering, may have
	      constrained the range of perspectives captured. Some respondents noted
	      that certain questions felt leading or that response options did not
	      fully reflect their views. As a result, nuanced or minority opinions
	      may be underrepresented, and the findings should be interpreted with
	      this limitation in mind. Future studies could incorporate more
	      open-ended items or cognitive pre-testing to better capture the
	      diversity of attitudes toward AI agents.
\end{enumerate}

\subsection{Future Research
	Directions}\label{future-research-directions}

This study suggests several promising avenues for future research:

\begin{enumerate}
	\def\labelenumi{\arabic{enumi}.}
	\item
	      \textbf{Longitudinal Studies}: Tracking how perceptions change as AI
	      agent capabilities evolve would provide valuable insights into
	      adaptation processes, particularly as agent frameworks mature from
	      lower TRLs to production-ready systems \cite{Wrona2023}.
	\item
	      \textbf{Comparative Analyses}: Examining differences in perceptions
	      across stakeholder groups (developers, managers, end users,
	      regulators) could inform targeted communication and implementation
	      strategies. \cite{LiuCHI2024} finding of significant cultural
	      differences in AI agent perceptions suggests that geographic and
	      cultural factors should be explicitly considered in future work.
	\item
	      \textbf{Organizational Case Studies}: Detailed examinations of
	      organizations that have successfully deployed agent systems could
	      identify best practices for overcoming the regulatory and compliance
	      barriers identified in our survey. Following \cite{Brown2025}
	      approach, studies could investigate how different feedback and
	      transparency mechanisms affect trust in autonomous agents within real
	      organizational contexts.
	\item
	      \textbf{Regulatory Framework Development}: Research on governance
	      models specifically adapted to the unique challenges of autonomous
	      agents could address the identified compliance barriers. This might
	      include exploring distributed responsibility models that align with
	      our respondents' preference for shared accountability among
	      developers, users, and the agent itself.
	\item
	      \textbf{Skill Transition Mapping}: Building on \cite{NoyZhang2023}
	      productivity findings, studies on how programming and knowledge work
	      skills are evolving in response to AI agents could inform educational
	      and workforce development initiatives. This could help prepare workers
	      for the supervisory and high-level design roles that our respondents
	      envisioned as the future of programming.
\end{enumerate}

\section{Conclusion}\label{conclusion}

This study provides empirical insights into how professionals perceive
autonomous AI agent systems---their capabilities, impact, governance
requirements, and ethical implications. The findings reveal a nuanced
landscape where most respondents acknowledged AI's growing impact on
programming while favoring collaborative models that maintain meaningful
human oversight.

The identification of three distinct respondent segments based on their patterns of responses highlights the
diversity of perspectives even within what is likely a technically
informed sample. These segments can help technology developers,
organizational leaders, and policymakers craft targeted strategies to
address specific concerns and leverage particular opportunities.

The prominence of regulatory concerns as a barrier to deployment (identified in Q3 responses)
reinforces recent findings by \cite{Wrona2023} regarding the
readiness gap between academic prototypes and production-grade systems,
emphasizing that autonomous AI agent adoption is not solely a technical
challenge but also a governance and compliance challenge. While our logistic regression model did not find regulatory concerns to be a statistically significant predictor of non-deployment when other factors were considered, its high ranking as a perceived barrier in the descriptive data warrants attention. Organizations
seeking to implement these technologies should invest in robust
frameworks for accountability, auditability, and oversight alongside
technical capabilities.

Our findings on human oversight preferences align with \cite{Hauptman2022} research showing that optimal agent autonomy depends on task
structure and work processes, suggesting that one-size-fits-all
implementation strategies are unlikely to succeed. Instead,
organizations should consider the nature of specific workflows and team
structures when determining appropriate levels of agent autonomy.

The productivity gains documented by \cite{NoyZhang2023} suggest that
autonomous agents could deliver substantial value if deployed
effectively, but our survey indicates that realizing these benefits will
require addressing significant trust and control concerns. \cite{Brown2025} work on real-time feedback mechanisms offers promising
directions for building user trust while maintaining human oversight.

As we enter an era where AI agents can increasingly operate with minimal
human supervision, finding the right balance between autonomy and
control, efficiency and safety, innovation and responsibility will be
crucial. This study suggests that stakeholders are already grappling
with these trade-offs, with most favoring approaches that preserve human
judgment while leveraging AI capabilities---a balanced perspective that
could guide responsible development and deployment of agent systems
across industries.

In a broader context, our results contribute to emerging cross-cultural
understandings of AI agent perception \citep{LiuCHI2024} by providing
detailed insights from a predominantly Western technical audience. This
adds an important dimension to the global conversation about AI agent
adoption and governance, highlighting specific concerns that may vary
across cultural and professional contexts.

Future work should address the limitations identified in our study,
particularly through larger, more diverse samples and longitudinal
designs that can track how perceptions evolve as AI agent capabilities
mature from current TRL levels to production-ready systems. The gap
between technical possibility and organizational adoption identified in
our study represents both a challenge and an opportunity for
researchers, developers, and organizational leaders working to shape the
future of autonomous AI agents.

\section{Acknowledgements}\label{acknowledgements}

We sincerely thank survey participants for generously sharing their time and insights. Special thanks to O'Reilly Media for organizing the event and the associated Discord group.

\section*{Data and Code Availability} 
The survey data, analysis scripts, and other materials supporting the findings of this study are openly available in the GitHub repository at \url{https://github.com/nkkko/agent-perceptions}.

\bibliographystyle{unsrtnat}
\bibliography{references}

\section{Appendix A: Survey Items}\label{appendix-a-survey-items}

\textbf{Q1: When do you think AI will be able to replace professional
	programmers?} - Never - creativity can't be automated - 5-10 years for
most programming tasks - Already happening for simple tasks - Machines
will write 99\% of code by 2030 - No opinion

\textbf{Q2: Which industry do you think will be most disrupted by
	autonomous AI agents in the next 5 years?} - Software development and IT
- Finance and banking - Healthcare and medicine - Customer service and
support - No opinion

\textbf{Q3: What is the biggest barrier to deploying autonomous agent
	systems in your organization?} - We ARE deploying them now -
Regulatory/compliance or technical readiness concerns - Fear of the
unknown - Technical leadership doesn't understand the potential - No
opinion

\textbf{Q4: Who should be responsible when an autonomous agent makes a
	critical mistake or decision?} - Developers who created the agent - End
users who deployed the agent - The agent itself (as a legal entity) -
All parties share responsibility - No opinion

\textbf{Q5: What concerns you most about autonomous AI agents?} - They
won't be capable enough to be useful - They'll be capable but
unpredictable/dangerous - Human skills will atrophy due to overreliance
- No opinion

\textbf{Q6: What's your preferred model for human-agent collaboration?}
- Agents should operate independently without oversight - Agents should
suggest but humans must approve all actions - Humans should have final
decision-making authority - Agents should operate within strict
guardrails but independently otherwise - No opinion

\textbf{Q7: Would you sacrifice control for efficiency?} - Never
sacrifice control regardless of efficiency - Sacrifice some control for
significant efficiency gains - Sacrifice significant control for maximum
efficiency - No opinion

\textbf{Q8: Do you trust the claims made by AI agent developers?} - They
generally represent capabilities accurately - They sometimes exaggerate
capabilities - They frequently misrepresent capabilities - No opinion

\textbf{Q9: How should society approach the ethical trade-offs with
	autonomous agents?} - Prioritize innovation even with some safety risks
- Prioritize safety even at the cost of innovation - Case-by-case
evaluation with expert oversight - No opinion

\textbf{Q10: What will be the role of human programmers in an era of
	code-generating AI?} - Focus on high-level design while machines handle
implementation - Become supervisors and reviewers of machine-generated
code - Transition to user experience and requirements gathering - Made
completely obsolete - No opinion

\section{Supplementary Figures \&
  Tables}\label{supplementary-figures-tables}

\begin{figure}[htbp]

	{\includegraphics[width=\textwidth, keepaspectratio]{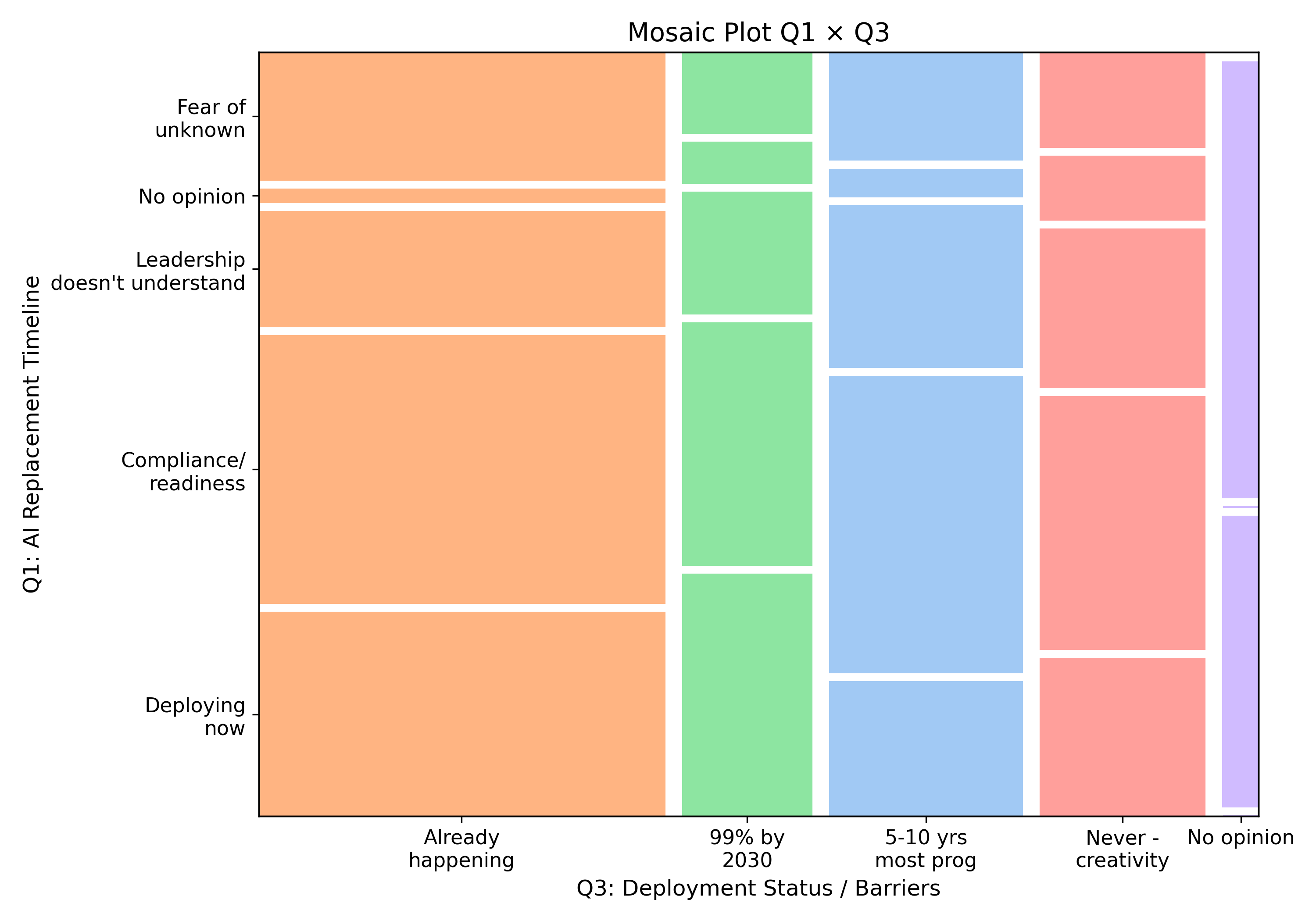}}

	\caption{Mosaic Plot Q1 × Q3 (Supplementary Figure S1).}\label{fig:q1xq3_mosaic}

\end{figure}%

\begin{table}[htbp]
	\caption{Summary of thematic clusters (0, 1, and 2) based on participant responses to qualitative questions (Q1--Q10) regarding autonomous agents. Each cell represents the dominant theme identified in that cluster for the corresponding question (Supplementary Table S1).}\label{table:kmodes_clusters}
	\centering
	\begin{tabularx}{\textwidth}{l>{\raggedright\arraybackslash}X >{\raggedright\arraybackslash}X >{\raggedright\arraybackslash}X}
		\toprule
		cluster & 0                                                          & 1                                                     & 2                                                          \\
		\midrule
		Q1      & Already happening for simple tasks                         & Never - creativity can't be automated                 & Already happening for simple tasks                         \\
		Q2      & Software development lifecycle                             & No opinion                                            & Knowledge work industry                                    \\
		Q3      & We ARE deploying them now                                  & Regulatory/compliance or technical readiness concerns & Regulatory/compliance or technical readiness concerns      \\
		Q4      & All decisions need human approval (defeating the purpose)  & The company deploying the agents                      & The company deploying the agents                           \\
		Q5      & Self-modification of their own architecture                & No opinion                                            & Self-modification of their own architecture                \\
		Q6      & Fluid/dynamic number based on task (emergent organization) & No opinion                                            & Fluid/dynamic number based on task (emergent organization) \\
		Q7      & Job security (including your own position)                 & No opinion                                            & Control (allow autonomous decision execution)              \\
		Q8      & The need for human oversight                               & No opinion                                            & Reasoning failures under uncertainty                       \\
		Q9      & Open source transparency                                   & No opinion                                            & Open source transparency                                   \\
		Q10     & Develop new types of work we can't yet imagine             & No opinion                                            & Provide oversight and creative direction                   \\
		\bottomrule
	\end{tabularx}
\end{table}

\end{document}